\documentclass[aps,showpacs,pra,notitlepage]{revtex4-1}

\usepackage[tbtags]{amsmath}
\usepackage{amssymb}
\usepackage{amsfonts}
\usepackage[dvips]{graphicx,color}

\begin{document}
\title{Simultaneous suppression of time and energy uncertainties in a single-photon frequency comb state}
\author{Changliang Ren}
\author{Holger F. Hofmann}
\email{hofmann@hiroshima-u.ac.jp} \affiliation{ Graduate School of
Advanced Sciences of Matter, Hiroshima University, Kagamiyama 1-3-1,
Higashi Hiroshima 739-8530, Japan} \affiliation{JST, CREST,
Sanbancho 5, Chiyoda-ku, Tokyo 102-0075, Japan}

\begin{abstract}
A single photon prepared in a time-energy state described by a frequency comb combines 
the extreme precision of energy defined by a single tooth of the comb with a high sensitivity to small shifts in time defined by the narrowness of a single pulse in the long sequence of pulses that describe the frequency comb state in the time domain. We show how this simultaneous suppression of time and energy uncertainties can be described by a separation of scales and compare this with the suppression of uncertainties in the two particle correlations of an entangled state. To illustrate the sensitivity of the frequency comb states to small shifts in time and frequency, we consider the Hong-Ou-Mandel dips observed in two-photon interference when both time- and frequency shifts between the input photons are varied. It is shown that the interference of two photons in equivalent frequency comb states results in a two dimensional Hong-Ou-Mandel dip that is narrow in both time and frequency, while the corresponding entangled photon pairs are only sensitive to temporal shifts. Frequency comb states thus represent a unique and different approach towards quantum operations beyond the uncertainty limit.
\end{abstract}

 \maketitle
\section{Introduction}

Although the time-energy degree of freedom is described by continuous variables, discretization is often implemented by time-bin encoding, since the encoding of qubit states
in a well-defined number of pulses centered around different times $t_n=T_n$ is more robust against decoherence effects, making time-bin encoding more suitable for long-distance quantum
communication \cite{Vallone,Cabello,Tittel,Sangouard,Berlin,Brendel,Marcikic,Gisin,Lamine}.
Importantly, time-bin encoding can also be used as a method of exploring energy-time entanglement when the temporal resolution of the detectors is limited, since the superposition of many time bins can define the frequency, and hence the energy of the photon, with arbitrarily high precision
\cite{Franson,Vallone,Cabello,Tittel,Sangouard,Berlin,Brendel,Marcikic,Gisin,Lamine}. Effectively, time-bin encoding means that the temporal resolution is low while the frequency resolution can be very high. If the same concept is applied in the frequency domain, one naturally arrives at the definition of a frequency comb, which is a coherent superposition of very narrow ``teeth'' centered around discrete frequencies $\omega_n=\Omega_n$ \cite{Udem99a,Udem99b,Jones,Holzwarth,Cundiff,Hall,Hansch}. Interestingly, a frequency comb can also be described by an equally spaced sequence of short pulses in the time domain. It therefore represents a combination of the high frequency resolution achieved by time-bin encoding with a high time resolution that is achieved by the phase-locking (or coherence) between the many frequencies of the comb. In the following, we will explore this simultaneous realization of high precision in time and in frequency in the context of single photons states. Specifically, we explore the similarities and the differences between this apparent violation of energy-time uncertainty with the well-known violation of uncertainty limits by two-photon entangled states. 

The rest of the paper is organized as follows. In section \ref{sec:comb} we introduce the fundamental concept of a single-photon frequency comb state and its non-classical properties.
In section \ref{sec:ent} we discuss the similarity between the single-photon frequency comb state and a corresponding two-photon time-energy entangled state and define an entangled state with time and frequency correlations that correspond to the shapes of the pulses and the frequency teeth of the comb state. In section \ref{sec:bunch}, we discuss the characterization of temporal sensitivity using two-photon interference and show how the Hong-Ou-Mandel dip of two independent comb state photons is slightly different from the Hong-Ou-Mandel dip of the corresponding two photon entangled state. In section \ref{sec:freqHOM}, we introduce a variable frequency shift in the two photon interference to obtain a frequency dependent Hong-Ou-Mandel dip. We show that this dip exhibits the high frequency sensitivity of the comb state, but not the frequency sensitivity of entanglement. In section \ref{sec:2DHOM}, we consider the combination of time and frequency shifts in the two-photon interference to produce a two dimensional Hong-Ou-Mandel dip. For frequency combs, this two dimensional dip can be much narrower than the uncertainty limit. It is therefore well suited to illustrate the non-classical properties of the frequency comb state. Section \ref{sec:concl} summarizes the results and concludes the paper.

\section{Separation of time and frequency scales in a single-photon frequency comb state}
\label{sec:comb}

A frequency comb is a coherent superposition of equally spaced optical
frequencies, corresponding to a similar superposition of equally
spaced short time pulses in the time domain. In the idealized limit
of infinitely long pulse trains, the bandwidth of each ``tooth'' of
the frequency comb approaches zero, corresponding to an arbitrarily
high precision in the frequency domain. At the same time, the length
of an individual pulse in time is only limited by the total
bandwidth covered by the comb, resulting in an arbitrarily high
precision in time. If only the vicinity of one of the discrete times
and frequencies is considered, it appears as if the frequency comb
could violate the Fourier limit of time and frequency. 

\begin{figure}
[ht]
\begin{center}
\includegraphics[width=0.35\textwidth]{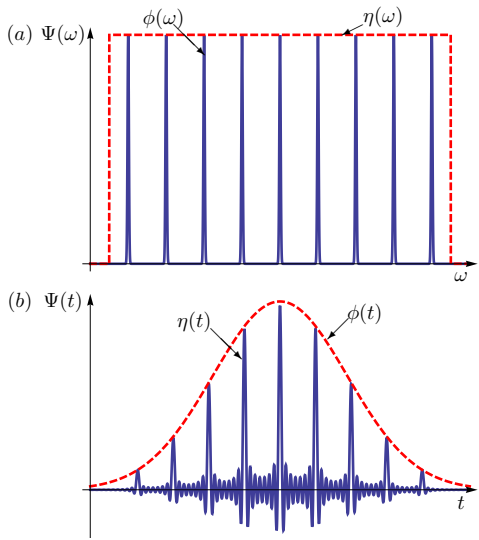}
\vspace{-10 pt}\caption{\label{comb} Illustration of the wavefunction of the frequency comb state in (a) frequency and (b) time. The example shown is defined by a rectangular spectral envelope and Gaussian spectral lines in (a), which result in the corresponding sinc-shaped pulses with a Gaussian temporal envelope in (b).
.}%
\end{center}
\end{figure}

Here, we consider the quantum state of a single photon in a frequency 
comb mode. This single photon state can be described by a coherent superposition 
of narrow frequency bands centered around discrete carrier frequencies separated 
by equal spacings of frequency difference $\Omega$. In the energy basis of the photon,
this state can be represented by the spectral envelope function $\eta(\omega)$ that characterizes the amplitudes of the discrete carrier frequencies, and a line shape 
function $\phi(\omega)$ around each carrier frequency that characterizes the precise shape of each tooth of the frequency comb. The single-photon frequency comb state can then be given by a superposition of energy eigenstates $\mid \omega \rangle$,
\begin{equation}
\label{eq:FCF}
\mid\Psi_c \rangle=\sqrt{\Omega}\int\eta(\omega)\sum_{n=-\infty
}^{\infty }\phi(\omega-n\Omega)\mid \omega\rangle \mathrm{d}\omega.
\end{equation}
As illustrated in Fig. \ref{comb} (a), the spectral envelope $\eta(\omega)$ determines the large scale features of the energy distribution and the line shape function $\phi(\omega)$ determines the microscopic details of the spectrum.  

To obtain the temporal shape of the frequency comb, it is convenient to express it in terms of the photon arrival time eigenstates $\mid t \rangle$. This representation is obtained by Fourier transformation of the spectral wavefunction. According to the convolution theorem, the temporal wavefunction is described by an envelope function $\phi(t)$ that is given by the Fourier transformation of the line shape function $\phi(\omega)$, and a pulse shape function $\eta(t)$ that is given by the Fourier transformation of the spectral envelope function $\eta(\omega)$, 
\begin{equation}
\label{eq:FCT} \mid\Psi_c\rangle=\sqrt{T}\int\phi(t)\sum_{n=-\infty
}^{\infty }\eta(t-nT)\mid t\rangle \mathrm{d}t.
\end{equation}
The time difference $T$ between the pulses is related to the frequency difference $\Omega$ of the comb by $T=2\pi/\Omega$. Fig. \ref{comb} (b) shows the Fourier transform of the frequency comb wavefunction shown in Fig. \ref{comb} (a). The rectangular spectral envelope $\eta(\omega)$ results in sinc-shaped pulses $\eta(t)$, whereas the Gaussian spectral line shape $\phi(\omega)$ results in a Gaussian envelope in time.

The characteristic properties of the frequency comb state are given by the time and frequency scales defined by the functions $\eta(\omega)$ and $\Phi(\omega)$, together with the frequency interval $\Omega$. In particular, the spectral width of the line shape $\Phi(\omega)$ can be identified with an energy uncertainty $\delta \omega_{\phi}$, while the spectral width of the envelope corresponds to a much larger energy uncertainty of $\delta \omega_{\eta}$. These two uncertainties define two different frequency scales which are related to the spectral separation $\Omega$ of the comb by
\begin{equation}
\label{eq:scale1}
\delta\omega_{\phi} < \Omega < \delta\omega_{\eta}.
\end{equation}
In the time domain, the uncertainty limit for each wavefunction requires that $\delta \omega_i \delta t_i \geq 1/2$. Therefore, the Fourier transform tends to reverse the order of the uncertainties, with the temporal uncertainty $\delta t_{\eta}$ of the pulse shape being minimal and the temporal uncertainty $\delta t_{\phi}$ being maximal,
\begin{equation}
\label{eq:scale2}
\delta t_{\eta} < T < \delta t_{\phi}.
\end{equation}
Importantly, the energy uncertainty $\delta \omega_{\phi}$ of the spectral lines and 
the time uncertainty $\delta t_{\eta}$ of each pulse are determined by completely different wavefunctions, so there is no uncertainty limit. Instead, there is a requirement that both are smaller than the distances $\Omega$ and $T$, which results in an {\it upper} limit for the uncertainty product given by
\begin{equation}\label{eq:crit}
\delta\omega_{\phi}\delta t_{\eta} <  2 \pi.
\end{equation}
Single photon frequency combs thus seem to exceed the uncertainty limit for energy and time. This is possible because the comb structure results in a separation of microscopic and macroscopic scales, as shown in Eqs. (\ref{eq:scale1}) and (\ref{eq:scale2}). Effectively, this scale separation within a single-photon state means that the microscopic details in time and in frequency commute with each other. Likewise, the macroscopic aspects described by the envelopes commute with each other, since they are also described by completely different wavefunctions. The uncertainty limit only applies to the relation between the microscopic energy and the macroscopic time represented by $\phi(\omega)$ and $\phi(t)$, and between the microscopic time and the macroscopic energy represented by $\eta(t)$ and $\eta(\omega)$. Frequency comb states may therefore shed some light on the relations between the macroscopic and the microscopic aspects of quantum systems, especially with regard to the possibility of macroscopic quantum effects \cite{Bru13,Lvo13,Wan13,Sek14}. 

\section{Analogy with time-energy entanglement}
\label{sec:ent}

As the discussion above shows, the separation between microscopic energy and microscopic time is achieved by defining two wavefunctions that appear to describe separate degrees of freedom. This situation is very similar to the separation of symmetric and anti-symmetric degrees of freedom in two particle entanglement. Specifically, the same suppression of uncertainties can be obtained in the correlations between two photons by replacing the microscopic energy distribution with the distribution over total energy, and the microscopic time distribution with the distribution over arrival time difference. 

Importantly, the wavefunction is now a genuine two particle wavefunction in the product space of two well separated degrees of freedom. The total two photon energy is given by the sum of two single photon energies, $\omega_1+\omega_2$. In down-conversion, this total energy can be described by a spectrally narrow function $\phi(\omega_1+\omega_2)$, so that a measurement of $\omega_2$ projects the state of photon 1 onto the line shape wavefunction $\phi(\omega_1)$ centered around a carrier frequency defined by $\omega_2$. The Fourier transformation of this wavefunction describes the average arrival time of the two photons, $(t_1+t_2)/2$. Using similar considerations for the anti-symmetric degree of freedom, the two photon state can be expressed in terms of a narrow line shape function $\phi(\omega)$ and a wide spectral envelope $\eta(\omega)$ as
\begin{equation}\label{eq:Estate1}
\mid\mathrm{E}_{1,2}\rangle=\int \int\phi(\omega_1+\omega_2)
\eta(\frac{\omega_1-\omega_2}{2})\mid\omega_1,\omega_2\rangle\mathrm{d}\omega_1\mathrm{d}\omega_2.
\end{equation}
The transformation to arrival time states results in the corresponding Fourier transforms of the functions $\phi(\omega)$ and $\eta(\omega)$, which then describe a long temporal envelope $\phi(t)$ and a short pulse shape $\eta(t)$. The temporal two photon wavefunction is given by
\begin{equation}\label{eq:Estate2}
\mid\mathrm{E}_{1,2}\rangle=\int \int\eta(t_1-t_2)
\phi(\frac{t_1+t_2}{2})\mid
t_1,t_2\rangle\mathrm{d}t_1\mathrm{d}t_2.
\end{equation}
Thus, a measurement of $t_2$ projects the state of photon 1 onto the pulse shape wavefunction $\eta(t_1)$ centered around a pulse time of $t_2$. Clearly, no uncertainty limit applies to the correlations in time and in frequency, since the frequency correlations are determined by the narrow spectral wavefunction $\phi(\omega)$ and the temporal correlations are determined independently by the short pulse $\eta(t)$.

In principle, we can observe very similar features in the spectral and temporal characteristics of photons in a frequency comb states and in the correlations between two photons in a time-energy entangled pair. In particular, the symmetric two photon wavefunctions $\phi$ in Eqs. (\ref{eq:Estate1}) and (\ref{eq:Estate2}) correspond to the combination of microscopic energy and macroscopic time in the frequency comb state, while the anti-symmetric wavefunctions $\eta$ correspond to the combinations of microscopic time and macroscopic energy. At first sight, the main difference is that there is no requirement similar to Eq.(\ref{eq:crit}), because entanglement does not depend on a clear separation of the time and frequency scales defining the functions $\phi$ and $\eta$. In practice, however, one would normally try to maximize the difference in scale, so that $\phi$ will effectively be microscopic in frequency, while $\eta$ will be microscopic in time. 

On closer inspection, a more important difference between entanglement and the frequency comb may be the need for two different symmetries to distinguish the two degrees of freedom in the two photon entangled state. Interestingly, there is a direct experimental demonstration of this difference: two photon interference can be used to explore the sensitivity to small shifts in time or in frequency and is also sensitive to the symmetry of the two particle wavefunction.  As we show in the following, the two-photon interference between photons in separate frequency comb states is equally sensitive to small shifts in time and in frequency, while the corresponding entangled states only show the sensitivities of the anti-symmetric degree of freedom, which is high with regard to time, but low with respect to frequency.

\section{Characterization of temporal sensitivity by two-photon interference}
\label{sec:bunch}

Since photon detection with high time resolution is difficult to implement using conventional detectors, it is often convenient to use the sensitivity of two photon interference to small delay times in order to investigate highly time resolved quantum states \cite{Ren,Hofmann}. In the present context, the time sensitivity of a single 
photon frequency comb state can be demonstrated by measuring the width of the Hong-Ou-Mandel dip that describes the dependence of coincidence counts on time delays between 
two photons in the same frequency comb state. It is then a straightforward matter to compare the result to the Hong-Ou-Mandel dip observed for two entangled photons defined 
by the same wavefunction.

\begin{figure}
[ht]
\begin{center}
\includegraphics[width=0.5\textwidth]{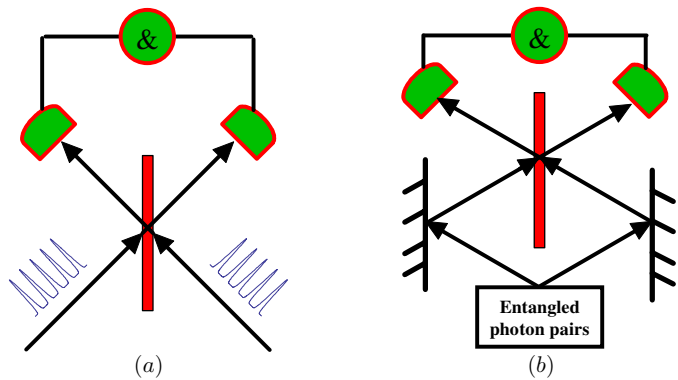}
\vspace{-5 pt}\caption{\label{HOM} Illustration of (a)
Hong-Ou-Mandel interference of two identical single-photon frequency comb states and (b) Hong-Ou-Mandel interference of a two-photon time-energy entangled state.}%
\end{center}
\end{figure}

Fig. \ref{HOM} illustrates the setups for Hong-Ou-Mandel interference for two photons in separate frequency comb states, and for a pair of entangled photons emitted from the same source. In both cases, the photons interfere at a $50:50$ beam splitter, and coincidence counts register the events where the photons exit from opposite sides of the beam splitter. Since the detectors do not resolve either time or frequency, the total coincidence count rate can be obtained by an integral over all possible arrival times at the detectors. If the two-photon input state is given by the temporal two-photon wavefunction $\psi_{p}(t_1,t_2)$ the coincidence counts of photons exiting at opposite sides of the beam
splitter is given by the integral
\begin{equation}\label{coin00}
C=\int\int\left\vert\psi_{p}(t_1,t_2)-\psi_{p}(t_2,t_1)\right\vert^{2}\mathrm{d}t_1\mathrm{d}t_2. 
\end{equation}
The effects of a time delay of $\Delta t$ applied only to the input photon entering from the left can be represented by shifting the first argument in the two functions, so that the time delay dependence of Hong-Ou-Mandel interference is generally described by
\begin{equation}\label{eq:cdt}
C(\Delta t)=\int\int\left\vert\psi_{p}(t_1-\Delta
t,t_2)-\psi_{p}(t_2-\Delta
t,t_1)\right\vert^{2}\mathrm{d}t_1\mathrm{d}t_2.
\end{equation}

We can now apply Eq.(\ref{eq:cdt}) to find the Hong-Ou-Mandel dip for the two photon interference between two separate single-photon frequency comb states as shown in Fig. \ref{HOM} (a). In this case, the input wavefunction is described by a product of two identical frequency comb wavefunctions $\Psi_c(t)=\langle t \mid \Psi_c \rangle$. Since the two-photon interference term is given by the overlap between the two input states \cite{Ren}, the time shift in Eq.(\ref{eq:cdt}) results in a convolution defined by the time delay $\Delta t$, so that the coincidence counts are given by
\begin{equation}\label{eq:combHOM}
C_{\mathrm{cs}}(\Delta t)=\frac{1}{2}-\frac{1}{2}\left\vert\int
\Psi_c(t)\Psi^{*}_c(t-\Delta
 t)\mathrm{d}t\right\vert^2.
\end{equation}
We can now solve the convolution integral by using the specific form of the 
frequency comb state shown in Eq.(\ref{eq:FCT}). The shape of the Hong-Ou-Mandel dip is then determined by an integral over a double sum,
\begin{equation}\label{eq:CC}
C_{\mathrm{cs}}(\Delta t)=\frac{1}{2}-\frac{1}{2}\left\vert T\int
\phi(t)\phi^{*}(t-\Delta t)\sum_{n',n''}\eta(t-n'T)\eta^{*}(t-\Delta
t-n''T)\mathrm{d}t\right\vert^2,
\end{equation}
where the sum runs over the pulse sequences entering the beam splitter from the 
right and from the left. 

The integration in Eq.(\ref{eq:CC}) can be simplified considerably by making use of the narrowness of the pulse shape function $\eta(t)$ in time. Since $\delta t_\eta \ll T$, the product of $\eta(t-n`T)$ and $\eta^{*}(t-\Delta t-n''T)$ will be negligibly small unless $t-n'T\ll T$ and $t-\Delta t-n''T\ll T$. We can therefore limit the integral to $t - n'T \ll T$ for time shifts selected by $\Delta t - m T \ll T$, where $m=n'-n''$. Since the envelope function $\phi(t)$ varies only slowly with time, we can approximate it using its value at the centers of the pulses described by $\eta(t)$. Using $t'=t-n'T$, we can then separate the sum over $n'$ from the integral over $t'$ to obtain
\begin{equation}\label{eq:Csep}
C_{\mathrm{cs}}(\Delta t) = \frac{1}{2}-\frac{1}{2}\left\vert
 \left(T\sum_{n'}\phi(n'T)\phi^{*}(n'T-\Delta
t) \right) \left(\sum_{m}\int\eta(t')\eta^{*}(t'-\Delta
t+mT)\mathrm{d}t'\right)\right\vert^2.
\end{equation}
Since the temporal envelope $\phi(t)$ varies slowly in time, the sum over $n'$ can be replaced by an integral and the result can be given in terms of the temporal autocorrelation functions of pulse shape $\eta(t)$ and envelope $\phi(t)$,
\begin{equation}\label{coin110}
C_{\mathrm{cs}}(\Delta t) = \frac{1}{2}-\frac{1}{2}\left\vert F_{\phi}(\Delta
t)\right\vert^2\left\vert\sum_{m}F_{\eta}(\Delta t+mT)\right\vert^2,
\end{equation}
where the two autocorrelation functions are defined as
\begin{equation}\label{F-phi}
F_{\phi}(\Delta t)=\int\phi(t)\phi^{*}(t-\Delta t)\mathrm{d}t
\end{equation}
and
\begin{equation}\label{F-eta}
F_{\eta}(\Delta t)=\int\eta(t)\eta^{*}(t-\Delta t)\mathrm{d}t.
\end{equation}
The width of these autocorrelation functions is closely related to 
the uncertainties of the corresponding wavefunctions. In particular, $F_{\eta}(\Delta t)$ will rapidly drop to zero for $\Delta t > \delta t_{\eta}$. Oppositely, we can assume that $F_{\phi}(\Delta t) \approx 1$ for $\Delta t\ll T$, since $\phi(t)$ changes only very little for such small time shifts. For $\Delta t\ll T$, it is therefore sufficient to consider only the pulse shape function $\eta(t)$, and the Hong-Ou-Mandel dip of the frequency comb state is given by 
\begin{equation}\label{eq:csdip}
C_{\mathrm{cs}}(\Delta t) =
\frac{1}{2}-\frac{1}{2}\left\vert F_{\eta}(\Delta t)\right\vert^2
\end{equation}
This Hong-Ou-Mandel dip is indistinguishable from the Hong-Ou-Mandel dip of two individual pulses defined by the wavefunction $\eta(t)$. Thus the high time resolution of the pulse shape function $\eta(t)$ can be observed in the narrowness of the Hong-Ou-Mandel dip observed for small time shifts of $\Delta t\ll T$.

Since it is a well-established experimental method to characterize the temporal correlations of down-converted photons by observing the Hong-Ou-Mandel dip using the setup shown in Fig. \ref{HOM} (b) \cite{Rubin, Grice, Giovannetti}, it is interesting to compare the result obtained from two frequency comb photons with the corresponding result from an entangled pair characterized by the same spectral and temporal correlations, as given in Eq.(\ref{eq:Estate1}). Using Eq.(\ref{eq:cdt}), we find that the Hong-Ou-Mandel dip for this entangled state is given by
\begin{equation}\label{eq:CE}
C_E(\Delta t)
=\frac{1}{2}-\frac{1}{2}\mathrm{Re}\left[\int \int \left(\phi(\frac{t_1+t_2-\Delta t}{2})
\phi^{*}(\frac{t_1+t_2-\Delta t}{2}))\right) \left(\eta(t_1-t_2-\Delta t)\eta^{*}(-t_1+t_2-\Delta t) \right) \mathrm{d}t_1 \mathrm{d}t_2 \right]
\end{equation}
The integral can be separated into an integral over the average time $(t_1+t_2)/2$, which is simply the normalization integral of the envelope function $\phi(t)$ and therefore gives a result of $1$, and an integral over the time difference $t_1-t_2$. Assuming that the pulse shape function $\eta(t)$ is symmetric in time, the Hong-Ou-Mandel dip can then be expressed by the same autocorrelation function used to describe the Hong-Ou-Mandel dip of the frequency comb. However, the result for the entangled photons has a slightly different form,
\begin{equation}\label{Edip}
C_E(\Delta t)=\frac{1}{2}-\frac{1}{2}\mathrm{Re}\left[F_{\eta}(2\Delta
t)\right].
\end{equation}
Although the width of this Hong-Ou-Mandel dip is determined by the same autocorrelation function that also determines it for the case of two photons in frequency comb states, it is worth noting that the observed shape is quite different from the one given in Eq.(\ref{eq:csdip}).
Specifically, the shape of the dip is given by the real part of the autocorrelation function, and not by the absolute value squared, and the factor of 2 in the argument indicates that the Autocorrelation function is only half as wide as the function seen in the Hong-Ou-Mandel dip of two independent frequency comb states. 

Due to the different shapes of the dips, it is not a straightforward matter to compare the widths. However, it should be noted that the square of a function always appears to be narrower than the function itself. For Gaussians, this results in a factor of $\sqrt{2}$, so the Hong-Ou-Mandel dip for frequency combs with Gaussian pulse shapes would appear to be only a factor of $\sqrt{2}$ wider than the Hong-Ou-Mandel dip of entangled photons with the same temporal correlations. Most importantly, the width of the Hong-Ou-Mandel dips is not directly related to the temporal uncertainties, but refers instead to the autocorrelation function of the pulse shape $\eta(t)$. Although this autocorrelation function appears in the Hong-Ou-Mandel dips of both the frequency comb states and the entangled states, the precise shape defined by the autocorrelation is quite different, indicating that the physics of two-photon interferences clearly distinguish the scale separation of the frequency comb from the separate degrees of freedom used in entanglement.

\section{Characterization of frequency sensitivity by two-photon interference}
\label{sec:freqHOM}

The main interest in the frequency comb state arises from the fact that it combines high time resolution with high resolution of photon energy. However, the conventional Hong-Ou-Mandel dip only shows the high time resolution associated with the pulse shape $\eta(t)$. To demonstrate the full symmetry between energy and time in the frequency comb state, it is therefore desirable to extend the two-photon interferences to frequency shifts between the two input photons. 

Although not as easy to implement as time delays, well-controlled shifts of frequency can in principle be realized by electro-optical modulation of the input field, or, for higher frequency shifts, by sum frequency generation in a non-linear medium. In the following, we will simply assume that the frequency can be shifted by a well-defined frequency difference $\Delta \omega$, just as the arrival time was shifted in the regular Hong-Ou-Mandel experiment. We can then obtain a spectral Hong-Ou-Mandel dip, where the coincidence counts can be determined from the spectral two-photon wavefunction $\psi_p(\omega_1,\omega_2)$ by
\begin{equation}\label{eq:freqC}
C_{\mathrm{cs}}(\Delta\omega)=\int\int\left\vert\psi_{p}(\omega_1-\Delta\omega,\omega_2)-\psi_{p}(\omega_2-\Delta\omega,\omega_1)\right\vert^{2}\mathrm{d}\omega_1\mathrm{d}\omega_2.
\end{equation}

For the frequency comb state, the spectral Hong-Ou-Mandel dip is obtained by simply exchanging the functions that describe the peaks and the envelopes in the temporal solution with the corresponding functions describing the spectral features. The result can be described in terms of the spectral autocorrelation functions of the line shape function $\phi(\omega)$ and the spectral envelope $\eta(\omega)$, 
\begin{equation}\label{eq:CSfreq}
C_{\mathrm{cs}}(\Delta\omega) = \frac{1}{2}-\frac{1}{2}\left\vert
F_{\eta}(\Delta\omega)
\right\vert^2\left\vert\sum_{m}F_{\phi}(\Delta\omega+m\Omega)\right\vert^2.
\end{equation}
The spectral autocorrelation functions are given by 
\begin{equation}\label{eq:Fw1}
F_{\eta}(\Delta\omega)=
\int\eta(\omega)\eta^{*}(\omega-\Delta\Omega)\mathrm{d}\omega
\end{equation}
and by
\begin{equation}\label{eq:Fw2}
F_{\phi}(\Delta\omega)=\int\phi(\omega)\phi^{*}(\omega-\Delta\omega)\mathrm{d}\omega.
\end{equation}
Since the spectral envelope $\eta(\omega)$ varies only slowly for frequency shifts of $\delta \omega \ll \Omega$, the shape of a spectral Hong-Ou-Mandel dip is given entirely by the autocorrelation function of the line shape $\phi(\omega)$,
\begin{equation}\label{eq:fdip}
C_{\mathrm{cs}}(\Delta\omega)= \frac{1}{2}-\frac{1}{2}\left\vert F_{\phi}(\Delta\omega)\right\vert^2.
\end{equation}
For the frequency comb states, the spectral Hong-Ou-Mandel dip will be as narrow as a single tooth of the frequency comb. Thus, two-photon interference can provide evidence of both the temporal and the spectral sensitivity of single photon frequency comb states, illustrating the possibilities of scale separation when encoding quantum information in continuous variable systems. 

Interestingly, the situation is quite different in the case of entangled pairs. Using the input state given in Eq.(\ref{eq:Estate2}), the spectral Hong-Ou-Mandel dip is given by
\begin{equation}\label{eq:Efreq}
C_E(\Delta \omega)
=\frac{1}{2}-\frac{1}{2}\mathrm{Re}\left[\int \int \left(\phi(\omega_1+\omega_2-\Delta \omega)
\phi^{*}(\omega_1+\omega_2-\Delta \omega))\right) \left(\eta(\frac{\omega_1-\omega_2-\Delta \omega}{2})\eta^{*}(\frac{-\omega_1+\omega_2-\Delta \omega}{2}) \right) \mathrm{d}t_1 \mathrm{d}t_2 \right]
\end{equation}
Here, the roles of $\eta$ and $\phi$ are not exchanged. As before, it is the function $\phi$ that can be eliminated by solving the normalization integral over the sum frequency $\omega_1+\omega_2$. Therefore, the coincidence counts are described by the autocorrelation function of the spectral envelope $\eta(\omega)$,
\begin{equation}\label{eq:flatspec}
C_E(\Delta\omega)=\frac{1}{2}-\frac{1}{2}\mathrm{Re}\big{[}F_{\eta}(\Delta\omega)\big{]}.
\end{equation}
Since the spectral envelope function $\eta(\omega)$ is much wider than $\Omega$, almost no coincidence counts will be observed for $\delta \omega \ll \Omega$. Because the Hong-Ou-Mandel dip of entangled photons is sensitive to the symmetry of the wavefunction, and not to the respective correlations, the high spectral precision of the line shape function $\phi(\omega)$ does not show up in the experimental two photon interference results. 

The absence of a spectral Hong-Ou-Mandel dip for entangled photons is a result of the symmetry of the spectrally squeezed degree of freedom. A similar effect was reported for the temporal Hong-Ou-Mandel dip by Giovanetti et al., who pointed out that two-photon entanglement with low time uncertainty in $t_1+t_2$ instead of $t_1-t_2$ results in an extremely wide temporal Hong-Ou-Mandel dip \cite{Giovannetti}. In that case, the Fourier transform of the anti-symmetric wavefunction is very narrow, which would result in a narrow spectral Hong-Ou-Mandel dip when applying frequency shifts instead of time delays. However, the use of symmetry to achieve a simultaneous suppression of uncertainties in two photon entangled states can not result in a two dimensional Hong-Ou-Mandel dip that is simultaneously narrow in both time and frequency, since the Hong-Ou-Mandel dip is only sensitive to one of the two degrees of freedom. We can conclude that the simultaneous suppression of uncertainties in the two dimensional Hong-Ou-Mandel dip is a characteristic feature of the frequency comb state that fundamentally distinguishes its non-classical properties from those of two photon entangled states. 

\section{Two-dimensional Hong-Ou-Mandel dip of two identical single-photon frequency comb states}
\label{sec:2DHOM}

Since the two-dimensional Hong-Ou-Mandel dip is a unique property of 
single photon frequency comb states, it may be of interest to consider 
the simultaneous application of a time delay $\Delta t$ and a frequency shift
$\Delta \omega$ in detail. If we delay the arrival time of one of the two input photons by $\Delta t$, and simultaneously shift the frequency of the other input
photon by $\Delta\omega$, the coincidence count rate can be described by
\begin{equation}\label{eq:cTF}
C(\Delta\omega,\Delta t)= \int\int\left\vert\psi_{p}(t_1-\Delta
t,t_2)e^{i t_2\Delta\omega }-\psi_{p}(t_2-\Delta t,t_1)e^{i
t_1\Delta\omega }\right\vert^{2}\mathrm{d}t_1\mathrm{d}t_2,
\end{equation}
where we have chosen to perform the analysis in the time representation, so that the frequency shift appears as a phase factor. In general, the simultaneous application of a temporal and a spectral shift cannot be separated and the result can be a complicated two-dimensional structure defined by the two-photon wavefunction $\psi_p(t_1,t_2)$ as a 
whole. However, the result can be simplified greatly if the input state is given by the product of two identical frequency comb states. 

Since we perform the analysis in the arrival time representation, the integral that determines the coincidence count rates is very similar to the one in Eq.(\ref{eq:CC}), except for a phase factor that represents the frequency shift,
\begin{equation}\label{eq:tfcomb}
C_{\mathrm{cs}}(\Delta \omega, \Delta t)=\frac{1}{2}-\frac{1}{2}\left\vert T\int
\phi(t)\phi^{*}(t-\Delta t) e^{i t\Delta\omega}
\sum_{n',n''}\eta(t-n'T)\eta^{*}(t-\Delta
t-n''T)\mathrm{d}t\right\vert^2.
\end{equation}
As before, we can make use of the fact that the pulse shapes $\eta(t)$ are very narrow in time to limit the integral to $t - n'T \ll T$ for time shifts selected by $\Delta t - m T \ll T$, where $m=n'-n''$. In general, the phase factor $\exp(i t \Delta\omega)$ should be included in the convolution integral of the pulse shape function $\eta(t)$, since a large value of $\Delta \omega$ causes a non-negligible variation of phase even within time intervals much smaller than $T$. However, we can simplify this relation if we consider only frequency shifts of $\Delta \omega \ll \Omega$, so that the phase change within an interval much smaller than $T$ is negligible. For consistency, we can also assume that $\Delta t \ll T$, so that only $m=0$ contributes to the Hong-Ou-Mandel dip. The sum and the integral can then be separated into
\begin{equation}\label{eq:tfsep}
C_{\mathrm{cs}}(\Delta \omega, \Delta t) = \frac{1}{2}-\frac{1}{2}\left\vert
 \left(T\sum_{n'}\phi(n'T)\phi^{*}(n'T) e^{i n' T \Delta\omega} \right) \left(\int\eta(t')\eta^{*}(t'-\Delta
t)\mathrm{d}t'\right)\right\vert^2.
\end{equation}
Here, the sum over $n'$ approximately describes the spectral autocorrelation function of the Fourier transformation of the temporal envelope $\phi(t)$, which is given by the line shape $\phi(\omega)$. It is therefore possible to express the two dimensional Hong-Ou-Mandel dip of the frequency comb state by
\begin{equation}\label{eq:2dHOM}
C_{\mathrm{cs}}(\Delta\omega,\Delta t) = \frac{1}{2}-\frac{1}{2}
\left\vert F_{\phi}(\Delta\omega)\right\vert^2\left\vert F_{\eta}(\Delta
t) \right\vert^2.
\end{equation}
Thus the two-photon interference of independent frequency comb states results in a two-dimensional Hong-Ou-Mandel dip that is simultaneously narrow in time and narrow in frequency. 

As we already know from the previous section, this result does not apply to the entangled state, since the two dimensional Hong-Ou-Mandel dip is only sensitive to the anti-symmetric wavefunction $\eta(t_1-t_2)$. Specifically, the coincidence counts for a simultaneous shift in time and frequency read
\begin{eqnarray}\label{eq:Eft}
\lefteqn{
C_E(\Delta\omega,\Delta
t)=}
\nonumber \\ &&
\frac{1}{2}-\frac{1}{2}\mathrm{Re}\left[\int\int \left(\phi(\frac{t_1+t_2-\Delta
t}{2})\phi^{*}(\frac{t_1+t_2-\Delta
t}{2})\right)\left( \eta(t_1-t_2-\Delta t)\eta^{*}(-t_1+t_2-\Delta
t)e^{-i (t_1-t_2)\Delta\omega }\right)\mathrm{d}t_1 \mathrm{d}t_2\right].
\end{eqnarray}
Since the integral over the average time $(t_1+t_2)/2$ is simply the normalization integral of $\phi(t)$, the shifts in time and in frequency act only on $\eta(t)$. Specifically, 
\begin{equation}\label{eq:Eeta}
C_E(\Delta\omega,\Delta
t)=\frac{1}{2}-\frac{1}{2}\mathrm{Re}\left[\int\eta(t)\eta^{*}(-t-2\Delta
t)e^{-i t\Delta\omega }\mathrm{d}t\right].
\end{equation}
In general, the two dimensional autocorrelation function of $\eta(t)$ determines the complete two dimensional Hong-Ou-Mandel dip for the entangled photon pairs. Since we are mostly interested in the sensitivity to small shifts in time and in frequency defined with relation to the time scale $T$ and the frequency scale $\Omega$, we can further simplify the result by noting that non-vanishing contributions to the integral are only obtained at $\Delta t\ll T$. In that region, the phase change caused by frequency shifts of $\Delta \omega \ll \Omega$ is always much smaller than $\Omega T = 2 \pi$ and can be neglected. It is therefore reasonable to approximate the two dimensional Hong-Ou-Mandel dip of the entangled pair by
\begin{equation}\label{coin-te}
C_E(\Delta\omega,\Delta t)\approx
\frac{1}{2}-\frac{1}{2}\mathrm{Re}(F_{\eta}(2\Delta t)).
\end{equation}
As discussed in section \ref{sec:freqHOM}, the reduced frequency uncertainty of the entangled state does not show up in two-photon interference. Therefore the two dimensional Hong-Ou-Mandel dip of entangled photon pairs is uncertainty limited, as opposed to the two dimensional Hong-Ou-Mandel dip for pairs of photons in frequency comb states. The narrowness of the two dimensional Hong-Ou-Mandel dip in both time and frequency is therefore a characteristic result of the suppression of uncertainties achieved by the separation of time and frequency scales in single photon frequency comb states.

\section{Conclusions}
\label{sec:concl}

Single photon frequency comb states combine the high temporal precision of short time pulses with the precise definition of photon energy of a narrow spectral line. Here, we have shown how the simultaneous suppression of uncertainty in time and in frequency can be
explained in terms of a separation of scales that separates the microscopic time and the microscopic frequency into entirely separate degrees of freedom, in the same way that two photon entanglement separates the symmetric and the anti-symmetric degree of freedom for the particle pair. In order to highlight the physical difference between the uncertainty suppression by scale separation and the apparent violation of uncertainty limits by energy-time entangled photon pairs, we have considered the sensitivities of two photon interference to small shifts in time and in frequency. Significantly, only the frequency comb states achieve simultaneous sensitivity to both time shifts and frequency shifts. This result indicates that single photon frequency comb states belong to a new class of non-classical states that obtains its non-classical features from a separation of scales, resulting in an effective commutativity of microscopic energy and microscopic time. Thus single photon frequency comb states are not only interesting for possible applications in new quantum protocols using the time-energy degree of freedom, but also for fundamental research into the relation between microscopic and macroscopic phenomena in quantum mechanics. 

\section*{Acknowledgments}
This work was supported by JSPS KAKENHI Grant Number 24540427.

%========================================================================================

\end{document}